\documentclass[Physsubmission, Phys]{SciPost}

\usepackage{mathtools}
\usepackage{appendix}
\usepackage{amsmath,graphicx,verbatim,epsfig}
\usepackage{amsfonts}
\usepackage{mathrsfs}  
\usepackage{epstopdf}
\usepackage[utf8]{inputenc}
\usepackage{amsmath,graphicx}
\usepackage{simplewick}
\usepackage{datetime}

\newcommand{\be}{\begin{equation}}
\newcommand{\ee}{\end{equation}}
\newcommand{\bea}{\begin{eqnarray}}
\newcommand{\eea}{\end{eqnarray}}
\newcommand{\balign}{\begin{align}}
\newcommand{\ealign}{\end{align}}

\newcommand{\st}{{\scriptscriptstyle T}}

\newcommand{\bg}{\begin{gather}}
\newcommand{\foma}{\end{gather}}

\newcommand{\noopsort}[1]{}

\def\slash{\rlap{/}}

\def\<{\langle}
\def\>{\rangle}

\def\({\left(}
\def\[{\left[}
\def\){\right)}
\def\]{\right]}

\def\Tr{\hbox{Tr}}

\newcommand{\ben}{\begin{eqnarray}}
\newcommand{\een}{\end{eqnarray}}

\newcommand{\bef}{\begin{figure}[htb]\centering}
\newcommand{\eef}{\end{figure}}

\hypersetup{
    colorlinks,
    linkcolor={red!50!black},
    citecolor={blue!50!black},
    urlcolor={blue!80!black}
}

\usepackage[bitstream-charter]{mathdesign}
\DeclareSymbolFont{usualmathcal}{OMS}{cmsy}{m}{n}
\DeclareSymbolFontAlphabet{\mathcal}{usualmathcal}

\def\AScom#1{{\bf  \textcolor{blue}{[AS: {#1}]}}}

\usepackage[normalem]{ulem}
\usepackage{slashed}

\begin{document}

\begin{center}{\Large \textbf{
Time reversal-odd effects in QCD and beyond\\
}}\end{center}

\begin{center}
Dylan Manna\textsuperscript{1$\star$},
Andrea Signori\textsuperscript{2,3,4} and
Christine Aidala\textsuperscript{1}
\end{center}

\begin{center}
{\bf 1} Department of Physics, University of Michigan, Ann Arbor, MI 48109, USA \\
{\bf 2} Dipartimento di Fisica, Universit\`a di Pavia, Pavia, Italy \\
{\bf 3} INFN, Sezione di Pavia, Pavia, Italy \\
{\bf 4} Theory Center, Jefferson Lab, Newport News, VA 23606, USA 

* dmanna@umich.edu
\end{center}

\begin{center}
\today
\end{center}


\definecolor{palegray}{gray}{0.95}
\begin{center}
\colorbox{palegray}{
  \begin{tabular}{rr}
  \begin{minipage}{0.1\textwidth}
    \includegraphics[width=22mm]{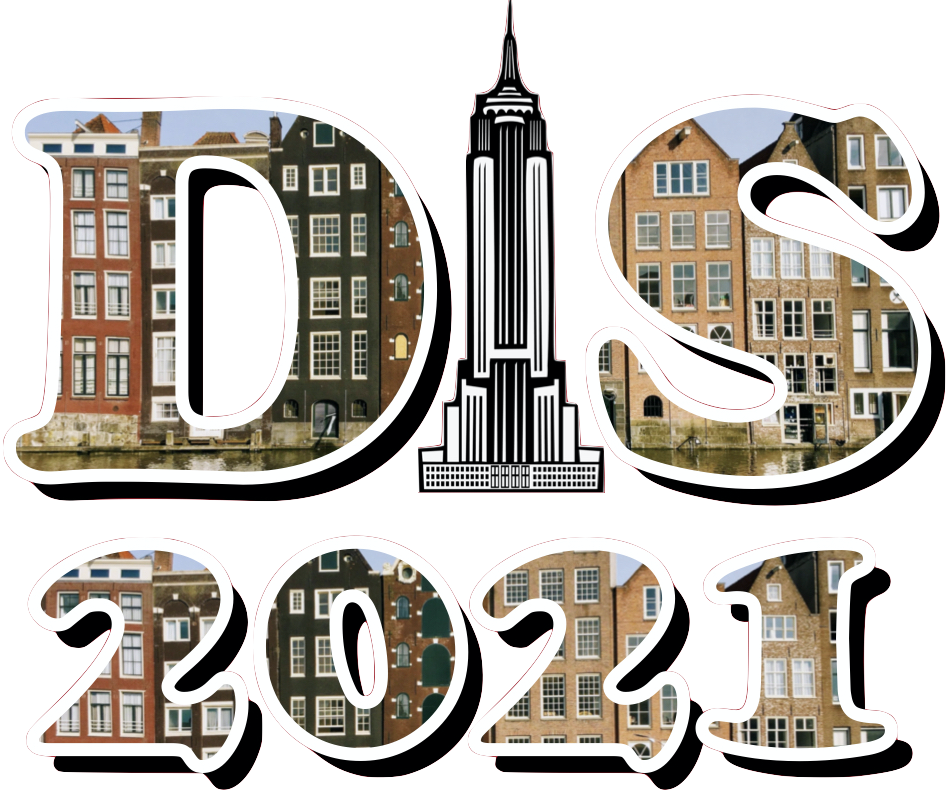}
  \end{minipage}
  &
  \begin{minipage}{0.75\textwidth}
    \begin{center}
    {\it Proceedings for the XXVIII International Workshop\\ on Deep-Inelastic Scattering and
Related Subjects,}\\
    {\it Stony Brook University, New York, USA, 12-16 April 2021} \\
    \doi{10.21468/SciPostPhysProc.?}\\
    \end{center}
  \end{minipage}
\end{tabular}
}
\end{center}

\section*{Abstract}
{\bf
In this paper we will describe a parallel between phenomenological and symmetry-based descriptions of the process dependence of the time-reversal-odd phenomena in QCD, such as the Sivers effect, with the goal of defining the essential elements that lead to such process dependence in QCD. 
This will then serve as a starting point to explore possible generalizations in a different gauge theory.
}


\section{Introduction}
\label{s:intro}
In this paper, we will first explore the general features of time-reversal-odd (T-odd) Transverse Momentum Dependent (TMD) Parton Distribution Functions (PDFs) and the associated process dependence in Quantum Chromodynamics (QCD). We will then elaborate on the possibility of accessing analogous effects in a different gauge theory, i.e., in Quantum Electrodynamics (QED). In order to understand the common physical and mathematical baseline for these effects, it is useful to briefly summarize the history of the problem.

In 1990, Sivers introduced a new quark distribution function sensitive to the transverse momentum of the parton and the transverse spin of the target, based on phenomenological arguments~\cite{Sivers:1989cc,Sivers:1990fh}.
In 1993, Collins argued~\cite{Collins:1992kk} that this type of function must be zero based on the combination of the parity ($P$) and the time-reversal ($T$) symmetries of QCD. 
In 2002, Brodsky, Hwang and Schmidt found~\cite{Brodsky:2002cx} a non-zero interference configuration potentially corresponding to a non-zero T-odd partonic distribution. In this context, a correlation between the transverse spin of the hadron and the partonic transverse momentum explicitly appeared in the formalism. 
Later in 2002, Collins amended his previous conclusions~\cite{Collins:2002kn} by the addition of Wilson lines in the definition of partonic distributions, and confirmed the potential existence of T-odd partonic distribution functions purely based on the interplay of the gauge symmetry and the time reversal symmetry of QCD.  

\section{Phenomenological approaches}
\label{s:pheno_based}

In 1976 at Argonne National Laboratory, there was a striking discovery of spin-dependent asymmetries in the momentum direction of charged pions produced via transversely polarized proton beams~\cite{Klem:1976ui}.  Since then, the study of spin asymmetries has rapidly evolved, both from the theoretical and the experimental point of view. 
In particular, the observations of spin asymmetries could not be successfully described within a simple collinear and leading-twist parton model. 
This triggered an investigation of hadron structure beyond this framework, and different (but complementary) mechanisms were proposed to account for spin asymmetries. 

The common feature of these mechanisms is a complex phase generated by an additional gluon exchange between the active parton and the remnant of the hadron target. 
Mathematically, this phase can be described with a dynamical higher-twist effect in collinear factorization or by the Wilson line in the TMD factorization formalism. 

\subsection{Sivers' approach}
\label{ss:Sivers}

In attempting to account for the large single spin asymmetries observed experimentally, in Ref.~\cite{Kane:1978nd} it was shown that a calculation purely based on collinear factorization in QCD cannot reproduce the observed effect, since at the partonic level the asymmetry would be largely suppressed being proportional to $\alpha_s \, m_q / \sqrt{s}$, where $\alpha_s$ is the coupling constant of QCD, $m_q$ the quark  mass, and $\sqrt{s}$ the energy in the center of mass. 

To attempt a qualitatively correct interpretation of these phenomena, Sivers introduced a non-perturbative partonic distribution sensitive to the transverse momentum of the quark as a generalization of collinear distributions~\cite{Sivers:1989cc,Sivers:1990fh}:
\begin{align}
\label{e:G_TMD}
& G_{a/h}(x;\mu) \longrightarrow G_{a/h}(x,k_T;\mu) \, , \\
\label{e:DeltaG}
& \Delta G_{a/h^\uparrow}(x, k_T ;\mu) = G_{a/h^\uparrow}(x, k_T ;\mu) - G_{a/h^\downarrow}(x, k_T ;\mu) \, ,
\end{align}
where $a$, $h$ are the parton and hadron flavors, $x$ and $k_T$ are the collinear momentum fraction and the transverse momentum of the quark, $\mu$ is the renormalization scale, and the arrow indicates the direction of the hadronic transverse spin. 
Estimates of the cross sections based on this function yield the correct order of magnitude for the observed asymmetries~\cite{Sivers:1989cc,Sivers:1990fh}. 
However, today we know that a description of the asymmetry for the specific process $pp^\uparrow \to h X$ in modern TMD factorization~\cite{Collins:2011zzd} is formally incorrect, and a collinear twist-3 formalism is more appropriate~\cite{Efremov:1981sh,Qiu:1991pp}. 



\subsection{Brodsky, Hwang, Schmidt's approach}
\label{ss:Brodksy}

\begin{figure}[htb]
\centering
\includegraphics[width=0.65\textwidth]{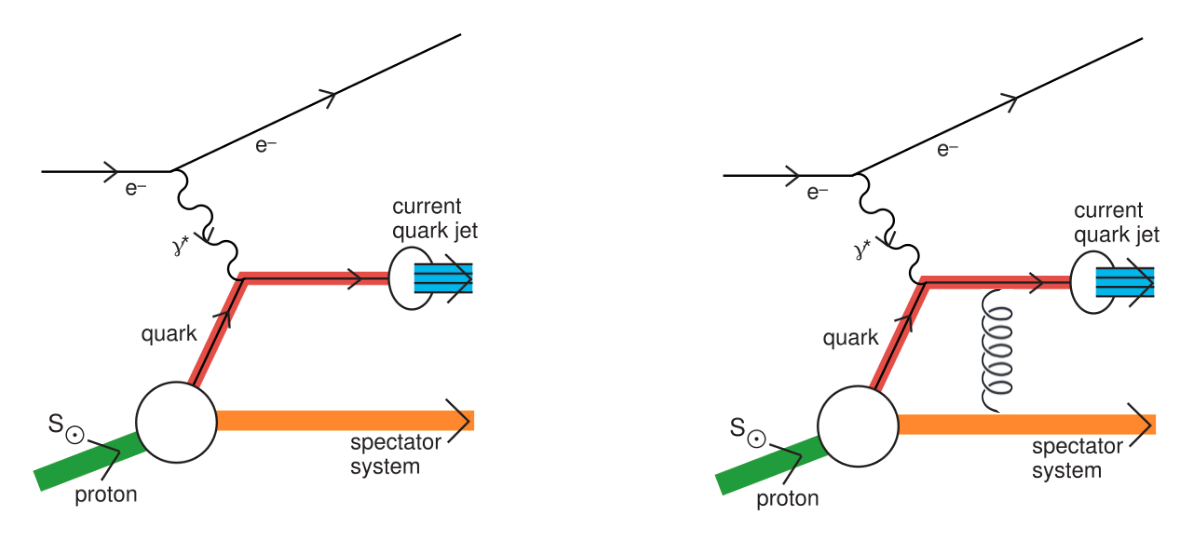}
\caption{SIDIS without and with gluon exchange. Original figure from Ref.~\cite{Brodsky:2002cx}.}
\label{f:brodsky}
\end{figure}

In 2002, Brodsky, Hwang, and Schmidt justify non-zero single spin asymmetries in QCD as produced by the interference between two amplitudes: one in which the outgoing quark and the hadronic remnant exchange a gluon and
one in which they do not~\cite{Brodsky:2002cx}. 
Fig.~\ref{f:brodsky} represents semi-inclusive deep inelastic scattering (SIDIS) without and with the aforementioned gluon exchange, whose interference generates a non-zero single spin asymmetry. 
In this context, the interference between these amplitudes is proportional to a spin-momentum correlation of the form:
\begin{equation}
\label{e:triple_prod}
i \vec S_p \cdot \vec q \times \vec p_q \, 
\end{equation}
with $\vec S_p$ the proton spin, $\vec p_q$ the quark momentum, and $q$ the photon momentum.

\section{Symmetry-based approach}
\label{s:symmetry_based}


The main criticism towards Sivers' arguments in favor of a non-zero partonic distribution function such as the one in Eq.~\eqref{e:DeltaG} came from Collins, who argued that such a time-reversal odd function should be zero based on the time reversal symmetry of QCD~\cite{Collins:1992kk}. 
These arguments 
were amended by Collins himself in 2002~\cite{Collins:2002kn} by using the correct gauge invariant definition for the quark correlation function. 
In the following we briefly outline the symmetry-based argument.

\begin{figure}[h!]
\centering
\begin{tabular}{ccc}
\includegraphics[width=0.25\textwidth]{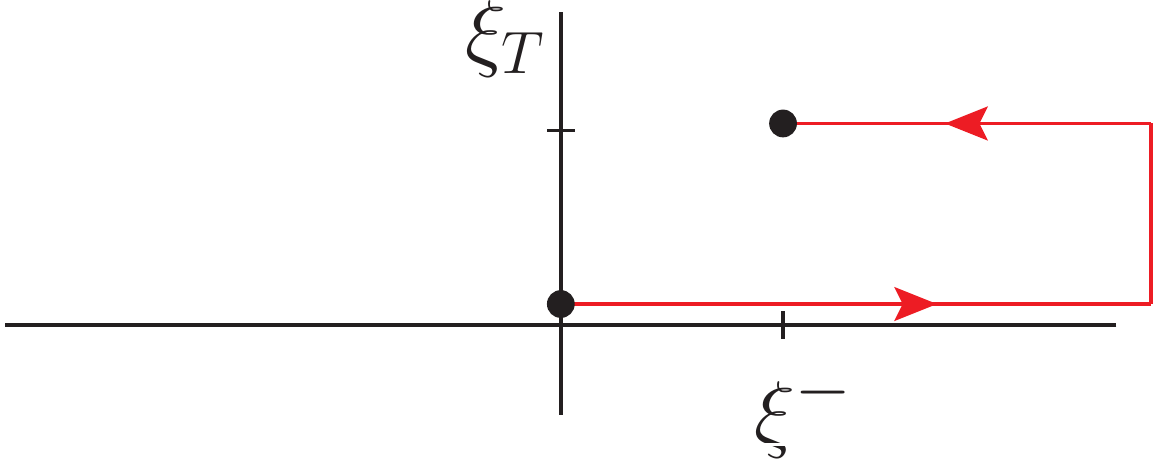}
& \hspace{2cm} &
\includegraphics[width=0.25\textwidth]{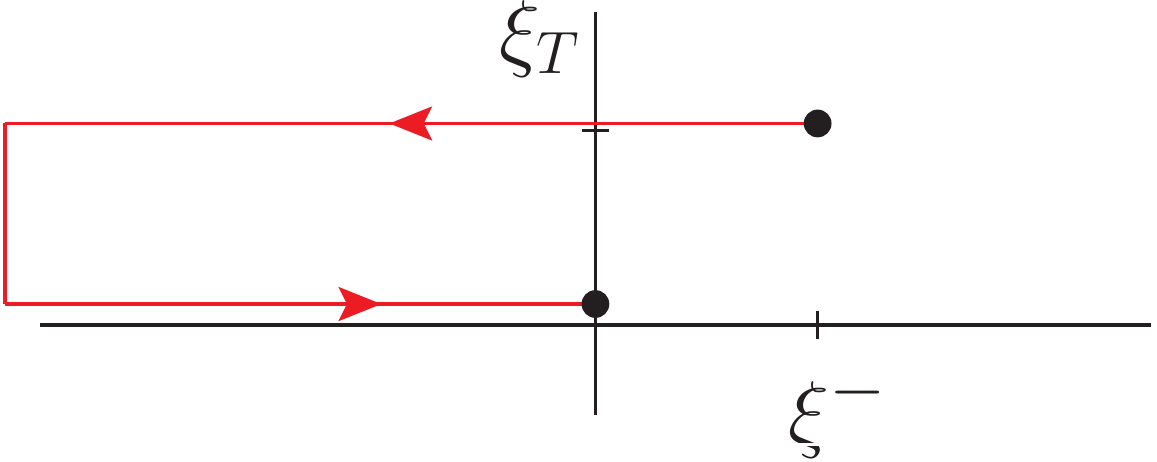}
\\
(a) & & (b)
\\ 
\end{tabular}
\caption{Representations of staple-like Wilson lines running from $0$ to $\xi$ in the hyperplane defined by the minus component and the transverse components of $\xi$: 
the future-pointing Wilson line $U_+$ (a) and the past-pointing Wilson line $U_-$ (b).}
\label{f:staples}
\end{figure}

Let us define the quark correlation function for a spin 1/2 hadron, including the Wilson line $U$ that guarantees the gauge invariance:
\begin{equation}
\Phi^{[U]}_{ij}(k,P,S) = \int\, \frac{d^4\xi}{(2\pi)^4} \ e^{i\,k\cdot \xi}  
\langle P,S \vert \overline \psi_j(-\xi/2)\, U(-\xi/2,\xi/2)\, \psi_i(\xi/2) \vert P,S \rangle \, ,
\label{e:qq_unint_corr_link}
\end{equation}
where $\psi$ is the quark field, $\xi$ is the non-locality in spacetime, $k$ is the quark four momentum, $P$ the hadronic four momentum, $S$ the covariant spin vector for the hadron, and the superscript $[U]$ represents the dependence on the path of the Wilson line. 
The path of the Wilson line is specified by the hard process in which the struck quark enters. For SIDIS the Wilson line is characterized by the future-pointing staple-like path described in Fig.~\ref{f:staples}~(a), whereas for Drell-Yan the path is a past-pointing staple (see Fig.~\ref{f:staples}~(b)).   
The time reversal symmetry of QCD implies that the unintegrated correlation function in Eq.~\eqref{e:qq_unint_corr_link} transforms in the following way:
\begin{eqnarray}
\label{e:qq_d_trev}
&\quad \Phi^{[\pm] *}(k;P,S) = i\gamma^1\gamma^3 \Phi^{[\mp]}(\tilde{k};\tilde{P},\tilde{S}) i \gamma^1\gamma^3 \ ,
\end{eqnarray}
where the superscripts $[\pm]$ refer to the staple-like Wilson line in Fig.~\ref{f:staples} and the tilde four vectors have opposite spatial components with respect to the non-tilde ones (e.g. $\tilde{k}^\mu = (k^0,-\vec{k})$).
Thus, the interplay between the time reversal symmetry and the Wilson line generates relations between the $\Phi^{[+]}$ and the $\Phi^{[-]}$ correlators, which allow one to define T-odd and T-even combinations of $\Phi^{[\pm]}$. 

Let us now introduce the transverse-momentum-dependent correlator for an unpolarized quark in a nucleon:
\begin{equation}
\label{e:TMD_corr_gp}
\Phi^{[U]}(x,\vec{k}_\st) = \frac{1}{2}\, f_1^{[U]}(x,k_\st^2)\, \slash{n}_+ 
					+ \frac{1}{2M}\, f_{1T}^{[U] \perp}(x,k_\st^2)\, \epsilon_\st^{\alpha\beta} {S_\st}_\alpha {k_\st}_\beta\, \slash{n}_+ \, ,
\end{equation}
where $f_1^{[U]}$ and $f_{1T}^{\perp [U]}$ are, respectively, the path-dependent unpolarized and Sivers TMD PDFs. 
%
The time-reversal relation in Eq.~\eqref{e:qq_d_trev} applied to the TMD correlator~\eqref{e:TMD_corr_gp} implies that:
\begin{equation}
\label{e:f1_f1Tp_proc_dep}
f_1^{[+]}(x,k_\st^2) = f_1^{[-]}(x,k_\st^2) \ , \quad \quad \quad 
f_{1T}^{\perp [+]}(x,k_\st^2) = -f_{1T}^{\perp [-]}(x,k_\st^2) \ .
\end{equation}
Eq.~\eqref{e:f1_f1Tp_proc_dep} means that the unpolarized TMD PDF $f_1$ is even under time-reversal transformation and universal, 
whereas the Sivers TMD PDF is odd under time-reversal transformation and is process dependent. 
This process dependence amounts to a sign-change between the SIDIS and Drell-Yan processes.  
Without the Wilson line operator $U$ in Eq.~\eqref{e:qq_unint_corr_link}, one would simply obtain $f_{1T}^{\perp} = -f_{1T}^{\perp} = 0$. This was indeed the original argument by Collins against T-odd distribution functions~\cite{Collins:1992kk} which was later corrected by introducing the proper Wilson lines~\cite{Collins:2002kn,Collins:2011zzd}.  
\section{QED vs QCD}
\label{s:QED_case}


Both QED and QCD potentially admit ``asymmetries'' which can be explained in terms of Wilson lines.
For example, relying on a double-slit experiment Aharonov and Bohm~\cite{Aharonov:1959fk} showed that the wave function of an electron moving around a solenoid experiences a phase shift as a result of the enclosed magnetic field.
The amplitude of the interference pattern is proportional to a phase of the form 
\begin{equation}
\label{e:AB_loop}
\exp \bigg\{ -i e \oint ds \cdot A(s) \bigg\} \, ,
\end{equation}
where $A(s)$ is the electromagnetic potential. 
The direction of the shift for the interference pattern changes sign depending on the direction of the magnetic field, which is similar to what happens in QCD with hadronic spin and the associated asymmetries.
Eq.~\eqref{e:AB_loop} is a QED Wilson loop, formally analogous to the non-abelian Wilson lines that mathematically justify the presence of T-odd partonic distributions in QCD and the emergence of single spin asymmetries therein. This analogy paves the way to the investigation of other possible QED observables sensitive to the interplay between the time-reversal and the gauge symmetry. 
Such experimental validation would complement and strengthen the existing QCD programs.





\section{Outlook} 
\label{s:outlook}

Single spin asymmetries in QCD and the associated T-odd partonic distributions can be investigated using two complementary approaches, one based on the symmetries of QCD and the other related to spin-momentum correlations (see e.g. Eq.~\eqref{e:triple_prod}). 
The T-odd effects are not peculiar to QCD but can be formally introduced as well in a different gauge theory with time-reversal invariance, for example QED. 
Finding an experimental confirmation of the process dependence of the T-odd distributions is one of the key-milestones of the QCD programs. 
We are planning to develop analogs of these QCD observables working in a different physics framework based on QED, for example atomic physics. 
This is motivated by the fact that 
from the theoretical point of view in QED one does not need to worry about non-perturbative physics and isolating this from the purely perturbative effects. 
Moreover, experimental measurements in atomic physics and QED are much more precise than in QCD.


This work is supported by the U.S. Department of Energy, Office of Science, Office of Nuclear Physics, contract no.~DE-SC0013393.  
AS acknowledges support from the European Commission through the Marie Sklodowska-Curie Action SQuHadron (grant agreement ID: 795475).  
JLab number: JLAB-THY-21-3476. 

\bibliography{t-odd_effects.bib}

\nolinenumbers

\end{document}